\documentstyle[12pt,psfig]{article}

\def\lp{\left(}
\def\rp{\right)}
\def\jpsi{$J/\psi$\,}
\def\psip{$\psi^{\prime}$\,}

\def\freezeout{FO\,}

\begin{document}

\begin{center}

{\large \bf Transverse Momentum Spectra of $J/\psi$ and $\psi^{\prime}$ Mesons from \\
Quark Gluon Plasma Hadronization at CERN SPS} 

\vspace*{0.5cm}

{\bf Kyrill A. Bugaev}

\vspace*{0.3cm}

Bogolyubov Institute for Theoretical Physics, Kiev, Ukraine\\

GSI, Darmstadt, Germany

\vspace*{0.1cm}

\hfill To O.L.

\end{center}

\begin{abstract}
A popular derivation of the apparent temperature  of the particle
transverse momentum spectra 
and its inconsistency are  considered. An improved  formula for the apparent temperature   is 
discussed. It is shown that recent results on transverse mass spectra of   
\jpsi and \psip mesons   support a hypothesis of statistical production
of charmonia at hadronization and suggest the early thermal freeze-out of these mesons.    
Using the apparent temperature formula  the collective
transverse velocity of the hadronizing quark gluon plasma is extracted  
to be $ \langle  v_T \rangle  \approx 0.2$.  
Predictions for transverse mass spectra of hidden and open charm mesons along with
bottomonium  at 
SPS and RHIC energies are discussed.
\end{abstract}


\noindent
{\bf 1. Statistical Production of Particles.} \jpsi\, suppression in  nuclear collisions was suggested 
by T. Matsui and H. Satz to be a signal of the deconfinement phase transition.
Nowadays there are several models which mainly differ from the original 
idea of the \jpsi suppression in the mechanism of  charmonia dissociation
either in quark gluon  or in nuclear media. 
Recently a principally different picture of the statistical  
charm production was suggested \cite{GG:99}.
The next breakthrough in this field, the statistical coalescence model, suggests \cite{BMGK} 
that
charmonium is generated by the coalescence of earlier produced $c$ and $\bar{c}$
quarks. 
However, important questions, what happens after hadronization and what are 
the transversal mass spectra 
of \jpsi\, and \psip\, mesons,  were not considered in \cite{BMGK}. 
Therefore, I will mainly concentrate on the problem of the \jpsi and \psip freeze-out
(\,\freezeout)
and will analyze their apparent temperature  according to \cite{BGG:01}. Then I will answer
the question, ``What can we learn from experimental
data about hadronization of quark gluon plasma (QGP)?''

The idealized concept of chemical (hadron multiplicities) and thermal (hadron momentum
distributions) freeze-outs enables one to interpret data on hadron production
in relativistic nucleus-nucleus (A+A) collisions.
The first experimental results on yields and transverse mass $(m_T = \sqrt{p_T^2 + m^2})$ spectra 
suggest the following scenario:
for the most abundant hadron species $(\pi, N, K, \Lambda)$ the chemical \freezeout,
which seems to coincide with the hadronization of the QGP, is followed by the 
thermal or kinetic \freezeout
occurring at a rather late stage of the A+A reaction.
Thus 
for the central Pb+Pb collisions at 158 A$\cdot$GeV the temperature of the chemical
\freezeout was extracted from the fit of the multiplicity data to be $T_H = 175 \pm 10$ MeV
\cite{All:99}.
The  thermal \freezeout parameters, i.e., temperature $T_F$ and averaged transversal velocity
$\langle v_T \rangle$,  for pions were determined
from the results of two pion correlations \cite{PionFR}
to be
quite different

\vspace*{-0.3cm} 

\begin{equation}\label{pifr}
T_F = 120 \pm 12 \,\,{\rm MeV},\quad \quad \langle v_T \rangle = 0.5 \pm 0.12\,\,.
\end{equation}

On the other hand  from hadronic cascade simulations it is known that thermal \freezeout\,   
of multistrange hadrons
($\phi, \Xi, \Omega$)
happens, probably, earlier than the kinetic \freezeout of pions. 
Now the main question is, ``When does  the \jpsi freeze-out occur? Late or early?'' 

The convolution of the transverse flow velocity of the 
matter element
with the thermal motion of hadrons in the rest frame of this element 
leads to a nearly  
exponential shape of final $m_T$ spectrum  
$\frac{1}{m_T} \cdot \frac{d~ N}{d~ m_T}
\approx C \cdot e^{\textstyle - m_T / T^* }$
with the apparent temperature (AT) $T^*$ which is defined
from the $p_T^2$ distribution at fixed longitudinal rapidity $y_p$ of the particle  as

\vspace*{-0.3cm} 

\begin{equation}\label{tdef}
T^* = 
 - \left[ 2 m_T \frac {d} {d p_T^2} \ln \lp \frac {dN} {d y_p\, d p_T^2} \rp \right]^{-1}\,\,.
\end{equation}
In the limit  $m_T >> T$ and $ \, p_T \rightarrow 0$  
the AT depends on the particle mass 
and $\langle v_T \rangle$
as follows 

\vspace*{-0.3cm} 

\begin{equation}\label{tapp}
 T^*  = T_F + \alpha \cdot m \cdot \langle v_T \rangle^2 \,\,.
\end{equation}
For the spherical fireball Eq. (\ref{tapp}) was first 
found
in Ref.  \cite{Zim}.
For the cylindrical geometry, surprisingly,  
there are   two answers  which differ by the value of the coefficient $\alpha$: 
there is a {\it naive result} $\alpha = \frac{1}{2}$ \cite{heiz:99} and 
{\it more elaborate} one $\alpha = \frac{2}{\pi}$ \cite{sin:99}. 
Therefore, before going further it is necessary to discuss the origin of
the difference between the results of Refs. \cite{heiz:99} and \cite{sin:99}.

\vspace*{0.3cm}

\noindent
{\bf 2. Particle Spectra and Apparent Temperature.}
%
The general  expression for the momentum distribution 
of the outgoing particle having the 4-vector of momentum
$p^\mu =$ $= (m_T\cosh y_p,\,\, p_T \cos(\phi_p),
\,\, p_T \sin(\phi_p), \,\, m_T \sinh y_p,)$ which is emitted from the 
fluid element having hydrodynamical 4-velocity $u^\mu = \gamma_T (\cosh y_L,\,\, v_T \cos(\phi_u),
\,\, v_T \sin(\phi_u), \sinh y_L,)$  of the arbitrary \freezeout hypersurface  
$\Sigma$ is given by the {\it cut-off distribution function}  
\cite{Bug:96, Bug:99a}

\vspace*{-0.3cm} 

\begin{equation}\label{cutdf}     
E \frac{d^3 N}{d p^3} = \int_\Sigma p^\mu\,d \Sigma_\mu~
\varphi \lp {\textstyle \frac{p^\nu u_\nu}{T} } \rp~
{ \Theta\lp p^\rho\, d \Sigma_\rho \rp }
\,\,,
\end{equation}
where the standard  notations  are assumed for 4-vectors 
($y_p$, $y_T$ and $y_L$  are particle longitudinal, fluid transversal and 
fluid longitudinal rapidities, respectively, and $\phi$ denotes the  corresponding polar angles),  
$\gamma_T = 1/\sqrt{1-v_T^2}$ is the relativistic
 $\gamma$-factor, $d \Sigma_\mu$
denotes the 4-vector of external normal to the \freezeout hypersurface $\Sigma$
and $\varphi$ is the one-particle phase space distribution function.
The $\Theta$-function in (\ref{cutdf}) is of  crucial importance
because it makes sure that only outgoing particles are counted 
 from both the time-
and space-like   
\freezeout hypersurfaces which are  defined, respectively,  
by the positive  sign $ds^2 > 0$ and negative
sign  $ds^2 < 0$ of the element square $ds^2 = dt^2 - dR^2 - dz^2 $ of the hypersurface 
(see \cite{Bug:96, Bug:99a} for detail). 

For further analysis I shall neglect the contributions coming from the time-like
parts of the \freezeout hypersurface and  then comment on general case. 
Now the  product $ p^\rho\, d \Sigma_\rho > 0$
is  positive for any momentum and  the {\it cut-off distribution} (\ref{cutdf})
automatically reproduces the famous Cooper-Frye result \cite{CF}. 
For the Boltzmann distribution $\varphi= \frac{g}{(2 \pi)^3} e^{\textstyle - \frac{p^\nu u_\nu}{T} }$
of particles 
emitted from the \freezeout hypersurface  $R^* = R^*(t, z)$
one obtains from (\ref{cutdf}) 

\vspace*{-0.5cm} 

\begin{eqnarray}\label{spectri}
\frac{d^2 N}{d y_p~d p_T^2} & = & \frac{g }{(2\pi)^2}\hspace*{-.1cm}\int_\Sigma~
dz~dt~ R^*~
e^{\textstyle - \frac{m_T\gamma_T \cosh (y_p-y_L)}{T}
  } m_T \cosh y_p~{ I_\phi} ({\textstyle \frac{p_T v_T \gamma_T}{T}}) \,\,,
\\
\label{iphi}
I_\phi ({\textstyle \frac{p_T v_T \gamma_T}{T} }) & = &
\int_{0}^{2\pi} d\phi_p~
\lp
{\textstyle \frac{\partial R^*}{\partial t} -
\tanh y_p \frac{\partial R^*}{\partial z} -
 \frac{p_T \cos (\phi_p) }{m_T \cosh y_p} } \rp
e^{\textstyle \frac{p_T  v_T \gamma_T \cos (\phi_p)}{T} }\,\,. 
\end{eqnarray}

\vspace*{-0.2cm} 

\noindent
Now it is clearly seen that 
the main contribution to the integral (\ref{iphi})
corresponds to the 
small angles $\phi_p$ between a 3-momentum of  particle and a 3-vector of the hydrodynamic velocity, 
whereas contributions coming from the large angles $\phi_p$ are suppressed exponentially. 

Neglecting the term with $\frac{p_T}{m_T}$  in Eq. (\ref{iphi}) for 
$p_T << m_T$ and expanding the Bessel function $I_0(x) \approx 1 + (0.5x)^2 $ 
for $  p_T v_T \gamma_T << T$, one finds  
from (\ref{tdef}) the following result for the AT 

\vspace*{-0.3cm} 

\begin{equation}\label{tnaive}
\frac{1}{T^*} \approx 
\left\langle\hspace*{-1.5mm}\left\langle 
\frac{ \gamma_T \cosh (y_p-y_L) - \frac{ m_T v_T^2 \gamma_T^2}{2\,T} }{T} 
\right\rangle\hspace*{-1.5mm}\right\rangle 
\approx 
\left\langle \frac{ \gamma_T  - \frac{ m_T v_T^2 \gamma_T^2}{2\,T} }{T} \right\rangle 
\end{equation} 
where the double averaging means a double integration over  
time and longitudinal  coordinate with the  weight function defined by
Eq. (\ref{spectri}). For  the longitudinal expansion     
depending on the rapidity $y_L$ only (e.g. Bjorken expansion)
the integration over $z$-coordinate   can be done  because  
for heavy particles the Boltzmann exponential behaves like a Kronecker $\delta$-function,
i.e., $e^{ - m_T \gamma_T \cosh (y_p-y_L) / T } \cong \delta(y_p-y_L) $.
This leads to a single averaging over the evolution time 
(last equality in Eq. (\ref{tnaive}) ) with a slightly
modified weight function.
The next approximation in (\ref{tnaive}) implies  
the  nonrelativistic transversal expansion and 
$m_T \rightarrow m$ 

\vspace*{-0.3cm}

\begin{equation}\label{tcond}
T^* \approx
T + \frac{m \left\langle v_T^2 \right\rangle }{2} 
 \approx
 T + \frac{m \left\langle v_T \right\rangle ^2 }{2}
\,\,, \quad {\rm for} \quad 
m v_T^2 \gamma_T << T\,\,, 
\end{equation}
where the last equality is fulfilled  within 10 \% for the linear dependence of the velocity
on the transversal radius.
Note, however, that a popular expression (\ref{tcond}) was obtained under the
condition which is hardly fulfilled in practice for heavy particles and, hence, 
Eq. (\ref{tcond}) cannot be established from Eq. (\ref{tnaive}).
Moreover,  for particles of $m_T \ge 1 $  GeV,  freezing out  
under conditions (\ref{pifr}), Eq. (\ref{tnaive})  leads to  negative 
contributions to the AT and even to  negative values of the AT! 
The latter follows from the fact that contributions coming from the small 
transversal radii (and, hence, small $v_T$) in (\ref{spectri}) are suppressed.

Thus an accurate examination of the derivation of Ref. \cite{heiz:99} 
shows that {\bf negative AT} can be seen, if the 
\freezeout  hypersurface is a space-like one.  
In Ref. \cite{Bug:99a} there are examples given that
{\bf negative AT} exist, if the  Cooper-Frye formula is applied to
the time-like hypersurfaces, whereas the cut-off formula (\ref{cutdf})
generates positive AT  in that case.
Discussion of the 
AT for  
the arbitrary hypersurface is
out of the scope of this talk. I mention only that 
negative AT not seen experimentally  
 may be an artifact of the Cooper-Frye formula.

In contrast an  improved derivation of Ref. \cite{sin:99} accounts for the
finiteness of the system and is partly free of the problems discussed above. 
In \cite{sin:99}  the  Cooper-Frye formula is modified by
an additional factor $\exp\{  - a (\cosh y_T -1) \}$ 
(with $ a = 1/ ( \left\langle R^* \right\rangle  v_T^{\prime} (0) )^2 $)
which in case of nonrelativistic transversal expansion reduces to 
$\exp\{  - \frac{R^{*\,2}}{2 \left\langle R^* \right\rangle ^2} \}$. 
Here $\left\langle R^* \right\rangle$ is the mean transverse radius
and $v_T^{\prime} (0)$ is the transverse velocity derivative at the center of the fireball.
The large radii contributions in (\ref{tnaive}) are suppressed due to  
such a factor and, hence, they do not lead to negative AT values. 
Further evaluation  ends  in (\ref{tapp}) with $\alpha = \frac{2}{\pi}$ which 
appears from the additional factor and reflects the cylindrical symmetry. 
Therefore, the latter is used below.

\vspace*{0.3cm}

\noindent
{\bf 3. Freeze-out of $J/\psi$ Meson.}
Recently it was found that 
the data on $J/\psi$ and $\psi^\prime$ yields in central Pb+Pb 
collisions at 158 A$\cdot$GeV are consistent with the  results of 
the statistical model for the typical value of
$T_H \cong 175$~MeV
extracted from light hadron systematics (see Refs. in \cite{GG:99, BGG:01}).
The hypothesis of statistical $J/\psi$ production at hadronization can
be further tested using data on $m_T$ spectra.
New  NA50  data  \cite{Jpsi}
on transverse mass spectra of $J/\psi$ mesons
in central Pb+Pb collisions at 158 A$\cdot$GeV confirm this expectation: the
spectrum is nearly exponential with  
$T^*(J/\psi) =245\pm 5$~MeV.
The measured $T^*(J/\psi)$ value is significantly smaller than 
expected one from 
Eq. (\ref{tapp}) 
for 
the pion \freezeout  parameters (\ref{pifr})  ($T^* \cong$ 610 MeV).

The `low' value of $T^*$ for $J/\psi$ suggests its rather early thermal
\freezeout.
The low interaction cross section of the $J/\psi$ meson with the
 most abundant hadrons \cite{cseci}
and its large mass lead to a very low probability of the $J/\psi$ rescattering on hadrons.
Following \cite{BGG:01}, it is assumed therefore that
the thermal \freezeout of $J/\psi$ coincides with the hadronization of QGP,
i.e., that the $J/\psi$ meson does not participate in the hadronic
rescattering after hadronization. It is, however, natural to expect that
there is a significant collective transverse flow of hadronizing QGP
developed at the early stage of partonic rescattering. Consequently the
AT of the $J/\psi$ meson as well as all other hadrons for
which chemical and thermal freeze--outs coincide with hadronization can be
expressed as:

\vspace*{-0.3cm} 

\begin{equation}  \label{slope1}
T^*_H (m) ~=~T_H~+~\frac{2}{\pi}~ \cdot m~ \cdot \langle v^H_T\rangle ^2~,
\end{equation}
where $\langle v^H_T\rangle$ is the mean transverse flow velocity of the QGP
at the hadronization. Assuming $T_H=175$~MeV and using the measured value of
$T^*(J/\psi) =245$ MeV one finds from Eq.(\ref{slope1}): $\langle v^H_T\rangle
\cong 0.19$. As expected the obtained transverse flow velocity of QGP at
hadronization is significantly smaller than the transverse flow velocity of
pions ($\approx 0.5$). The linear $m$--dependence of $T^*_H$ (\ref{slope1})
is shown in Fig. 1 by the lower solid line. Within the approach discussed
here, Eq.~(\ref{slope1}) can be used to obtain a next estimate of the lower
limit of the measured AT for all hadrons. In fact the
values of the parameter $T^*$ for all light hadrons are higher than $T^*_H(m)$. 
The recent results \cite{Jpsi} on
the $m_T$ spectra of the $\psi^{\prime}$ meson
indicate that $T^*(\psi^{\prime}) \approx T^*_H(\psi^{\prime}) = 258 \pm 5$
MeV, which suggests that  the $\psi^{\prime}$ meson (like $J/\psi$) does
not participate in the hadronic rescattering either.

One may expect that the thermal \freezeout may coincide with hadronization
also for $D$ and $\Upsilon$  mesons. Under this assumption one can calculate the value of the
apparent temperature for those mesons : $T^*(D) \cong T^*_H(D) \cong 217$ ~MeV
and $T^*(\Upsilon) \cong T^*_H(\Upsilon) \cong 392 $ ~MeV, respectively.

For the production of open and hidden charm particles
in A+A collisions at RHIC energies one can expect stronger transverse
collective flow effects than at the SPS. This will lead to a linear mass
dependence (\ref{slope1}) of the AT with approximately
the same value of $T_H\cong 175$~MeV, but with a larger value of $\langle
v_{T}^{H}\rangle $. A recent analysis \cite{Redl:01} 
of the RHIC data leads to the above
value $T_H\cong 175$~MeV of the chemical \freezeout temperature. 

Since  the RHIC data  for $J/\psi$ mesons are not    
measured yet, 
we use the hydrodynamic calculations of Ref.~\cite{Dum:99} which predict
the value of $\langle v_{T}^{H}\rangle \cong 0.30$ at the hadronization in
Au+Au collisions at RHIC. This leads to an increase of the AT of
charmed hadrons at RHIC in comparison to those values at SPS, e.g., $
T^*(J/\psi) \cong T_H^*(J/\psi)\cong 350$~MeV and $T^*(D) \cong T^*_H(D)
\cong 280$~MeV.
If bottomonium experiences the same \freezeout conditions at RHIC like $J/\psi$ meson,
then its apparent temperature will be about 
$T^*(\Upsilon) \cong T^*_H(\Upsilon) \cong 720 $ ~MeV. 

In summary, recent results on transverse mass spectra of $J/\psi$ and $%
\psi^{\prime}$ mesons in central Pb+Pb collisions at 158 A$\cdot$GeV are
considered. It is shown that these data support the hypothesis of the
statistical production of charmonia at hadronization and suggest a
simultaneous  hadronization and the thermal \freezeout for $J/\psi$ and $%
\psi^{\prime}$ mesons. Based on this approach the collective transverse
velocity of hadronizing quark gluon plasma is estimated to be $\langle v^h_T
\rangle \approx 0.2 $. Predictions for transverse mass spectra of hidden and
open charm along with hidden bottom
mesons at SPS and RHIC are discussed.
 
 \vspace*{0.3cm}


\noindent
{\bf Acknowledgments.}
I am thankful to M. Ga\'zdzicki, M. I. Gorenstein, P. T. Reuter 
and D. H. Rischke for valuable comments.
The financial support of the NATO Linkage
Grant PST.CLG.976950 is acknowledged.
The research described in this
publication was made possible in part by Award No. UP1-2119 of the U.S.
Civilian Research \& Development Foundation for the Independent States of
the Former Soviet Union (CRDF).



 \begin{figure}
\mbox{\hspace*{2.5cm}\psfig{figure=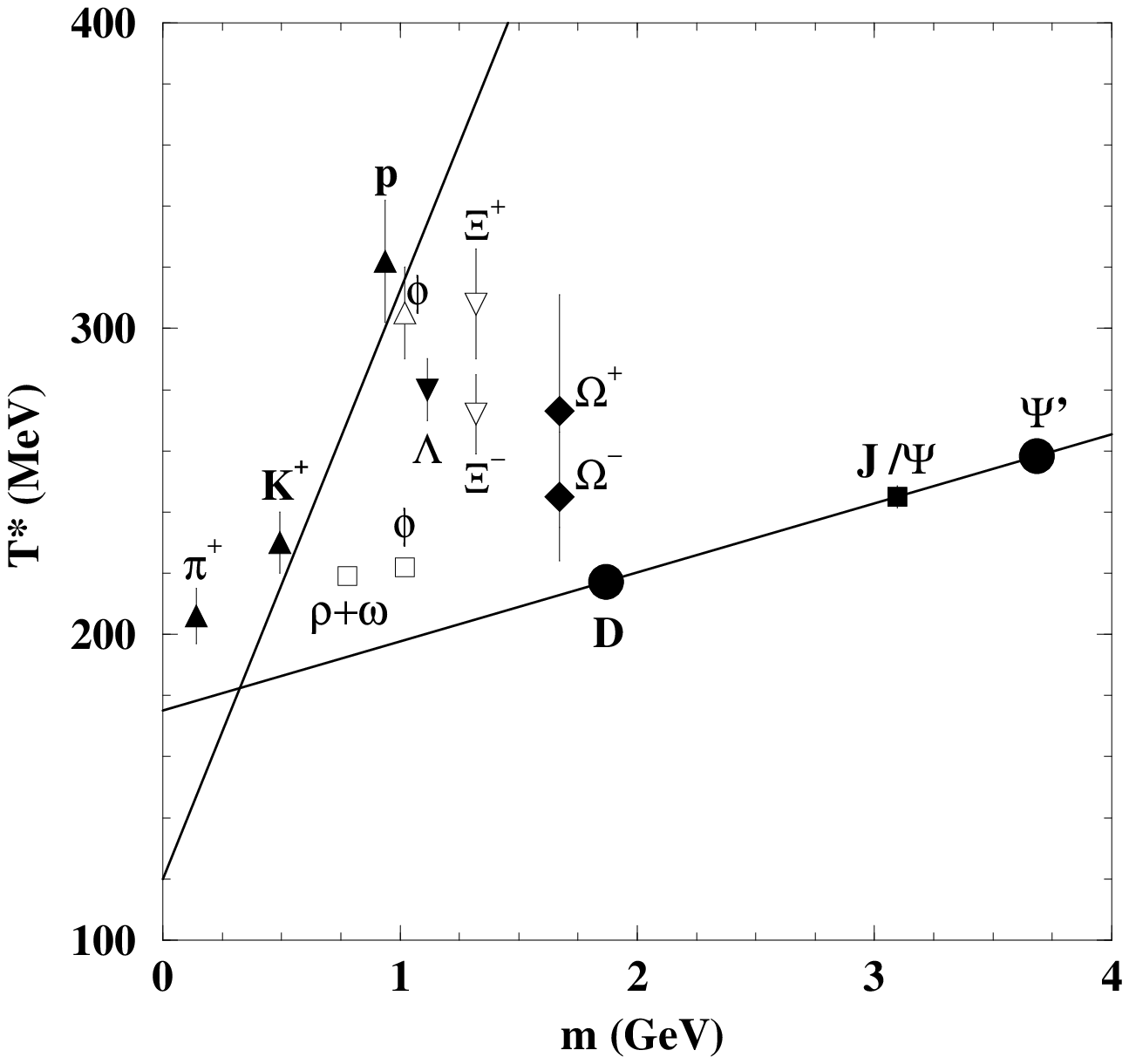,height=7.5cm,width=7.5cm}}

 \vspace*{-1.5cm}
  
  \noindent
  \caption{\label{fig:one}
  The inverse slope parameter as a function of the particle mass for
  central Pb+Pb collisions at 158 A$\cdot$GeV. The results 
  are compiled from: 
  \cite{trUF}
  (filled $\bigtriangleup$ ), 
  \cite{sqO}
  (open $\Box$),
  \cite{trUO}
  (open $\bigtriangleup$),
  \cite{trDF}
  (filled $\bigtriangledown$), 
  \cite{trDO}
  (open $\bigtriangledown$), 
  \cite{diam} (filled $\Diamond$), 
  \cite{Jpsi}
  (filled $\Box$).
  The filled circles are predictions of the model for $\psi^\prime$ and D
  mesons. The upper solid line  is given by Eq.~(3) with $\alpha = \frac{2}{\pi}$ 
  for the \freezeout
  parameters (\ref{pifr}). 
  The lower solid line is given by Eq.~(\ref{slope1}) with  $T_H=175$~MeV and
  $\langle v_T^H \rangle = 0.19$
  which correspond to the QGP hadronization.
  }
  \end{figure}


\def\plb{{\it Phys. Lett.} {\bf B}\,}
\def\prl{{\it Phys. Rev. Lett.\,}}
\def\jpg{{\it J. Phys. } {\bf G}\,}
\def\prc{{\it Phys. Rev.} {\bf C}\,}
\def\epj{{\it Eur. Phys. J.} \,}
\def\np#1{{\it Nucl. Phys.} {\bf #1}}
\def\jp#1{{\it J. of Phys.} {\bf #1}}
\def\preprint#1{{\it Preprint} {\bf #1}}
\def\hip#1{{\it Heavy Ion Phys.} {\bf #1}}


\small{

}

\end{document}